%% file: chapterCorrelations.tex
\pdfoutput=1
\documentclass[%
aps,amsmath,amssymb,floatfix,pra,notitlepage,superscriptaddress,longbibliography,nofootinbib,10pt
]{revtex4-1}

\usepackage{graphicx}
\usepackage[paperwidth=210mm,paperheight=297mm,centering,hmargin=4.0cm,vmargin=4.5cm]{geometry}

\makeindex

\include{commands}

\makeatletter
\newcommand{\manuallabel}[2]{\def\@currentlabel{#2}\label{#1}}
\makeatother

\begin{document}




\title{Quantum Noise Correlation Experiments with Ultracold Atoms}

\author{Simon F\"olling}


\affiliation{LMU M\"unchen, Fakult\"at f\"ur Physik, Schellingstr. 4, 80799 M\"unchen \\
Max-Planck-Institut f\"ur Quantenoptik, Hans-Kopfermann-Str. 1, 85748 Garching, Germany\\
Simon.Foelling@lmu.de} 

\begin{abstract}
Noise correlation analysis is a detection tool for spatial structures and spatial correlations in the in-trap density distribution of ultracold atoms. In this chapter, we discuss the implementation, properties and limitations of the method applied to ensembles of ultracold atoms in optical lattices, and describe some instances where it has been applied.
\vspace{0.7cm}

\noindent \emph{To appear as Chapter 8 in "Quantum gas experiments - exploring many-body states," P. T\"orm\"a, K. Sengstock, eds. (Imperial College Press, 2014)}

\end{abstract}

\maketitle   

\tableofcontents

\manuallabel{chapterTormaSengstock}{1}
\manuallabel{chapterKollath}{3}
\manuallabel{chapterKokkelmans}{4}
\manuallabel{chapterSengstock}{5}
\manuallabel{chapterChin}{6}
\manuallabel{chapterWeitenberg}{7}
\manuallabel{chapterFolling}{8}
\manuallabel{chapterParish}{9}
\manuallabel{chapterTorma}{10}
\manuallabel{Rabi}{10.2}
\manuallabel{chapterTarruell}{11}
\manuallabel{chapterKohl}{12}
\manuallabel{chapterSantos}{13}
\manuallabel{chapterPfau}{14}

\manuallabel{sec:Weitenberg_ParityProjection}{7.2.3}
\manuallabel{sec:Weitenberg_ShellStructure}{7.3.1}
\manuallabel{sec:imagingYYZ}{2.5.5}
\manuallabel{eq:TorontoOD}{2.20}
\manuallabel{sec:sequenceYYZ}{2.5}

\section{Noise and correlations in cold atoms\label{sec:sf_IntroNoiseCorrel}}
Experiments with ultracold atoms in optical lattices have attracted a lot of interest due to the ability to realize strongly interacting, strongly correlated many-body quantum states. Such strongly correlated states, however, are typically characterized by a loss or suppression of first-order coherence between lattice sites even for degenerate ensembles of Bosons derived from a coherent Bose-Einstein condensate (BEC)\cite{greiner02a,gerbier05a,gerbier05b,gerbier07c}. 
This has a profound impact on the measurements done on such ensembles, when using usual techniques of imaging after ballistic expansion of the atom cloud from the trap location. If interactions are negligible during expansion, such a measurement corresponds to a density measurement in momentum space, as opposed to the real-space distribution in the trap, as outlined in chapter\ref{sec:imagingYYZ}.

Because no fixed phase relation exists between the atoms released from different sites inside the lattice in the strongly correlated state, no interference appears in the expectation value between the wave functions from different sites. The signal obtained from a lattice filled with $N$ atoms therefore is the same as the expectation value of the signal from a single atom on a single site multiplied with $N$, without any information about the many-body state prior to release.

\begin{figure}
	\centering
		\includegraphics[width=1.0\textwidth]{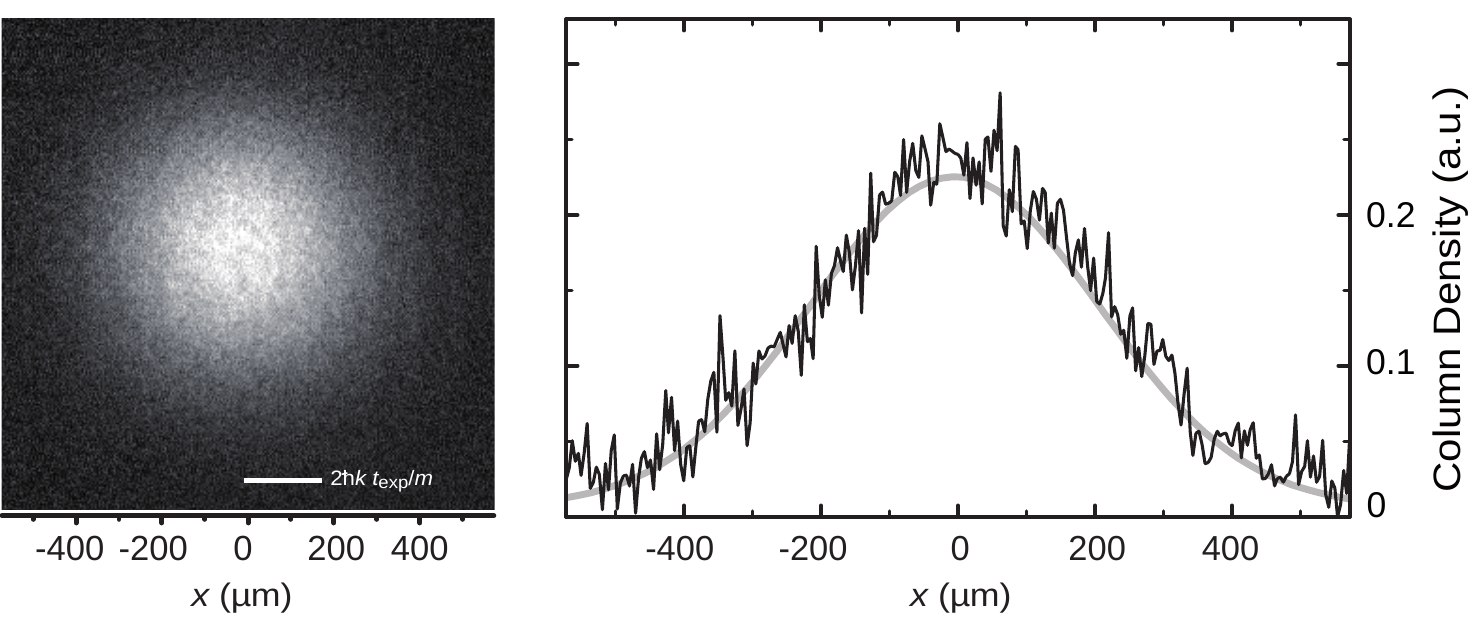}
	\caption{(a) Typical absorption image obtained for the case of a Rubidium Mott insulator in a deep lattice generated with an 840\,nm laser. A horizontal profile along the central line of the image is shown in part (b), together with a Gaussian fit (gray line). The column density in the center is approximately 30 atoms per pixel. The noise on the figure is dominated by the shot noise of this density as resolved by the camera system (figure adapted from \cite{foelling05}, Nature 434, 481, (2005)) }
	\label{fig:sf_GaussianImage}
\end{figure}

Such a distribution is shown in figure \ref{fig:sf_GaussianImage}, which illustrates that the density distribution does not contain much information --- in this case, it is extremely well described by a simple Gaussian, which corresponds to the expectation value for this trap.
Noise correlation analysis has been proposed\cite{altman04,grondalski99,kolovsky04,bach04} as one way of obtaining meaningful information about the state of the ensemble in the lattice prior to release anyway, by finding correlations in the fluctuations of the recorded signal, and thus in the deviations from the expectation value. As the particle number in a given region, such as that corresponding to a pixel in the image, is discrete, it is natural to assume that this number will fluctuate around the expectation value with a Poissonian distribution, with a scaling of $\sqrt n$, if $n$ is the number of atoms detected in that region. In figure \ref{fig:sf_GaussianImage} for example, where the measured column density of an expanded atom cloud is shown, this noise is clearly visible. In this section, we will describe why correlations exist in the noise, and how they are related to the ensemble in the trap prior to release.

\subsection{Origin of correlations between atom pairs\label{sec:sf_OriginOfNoiseCorrel}}
So, how do the fluctuations of the density signal contain information about the state of the ensemble even if the expectation value of the density does not? More fundamentally, why is there a correlation relation between different locations in space after atoms have been released even if the atoms are completely unrelated prior to their release? In the following, we will illustrate the origin of such correlations in the simple picture of two separate sources of particles.

\begin{figure}
	\centering
		\includegraphics[width=1.0\textwidth]{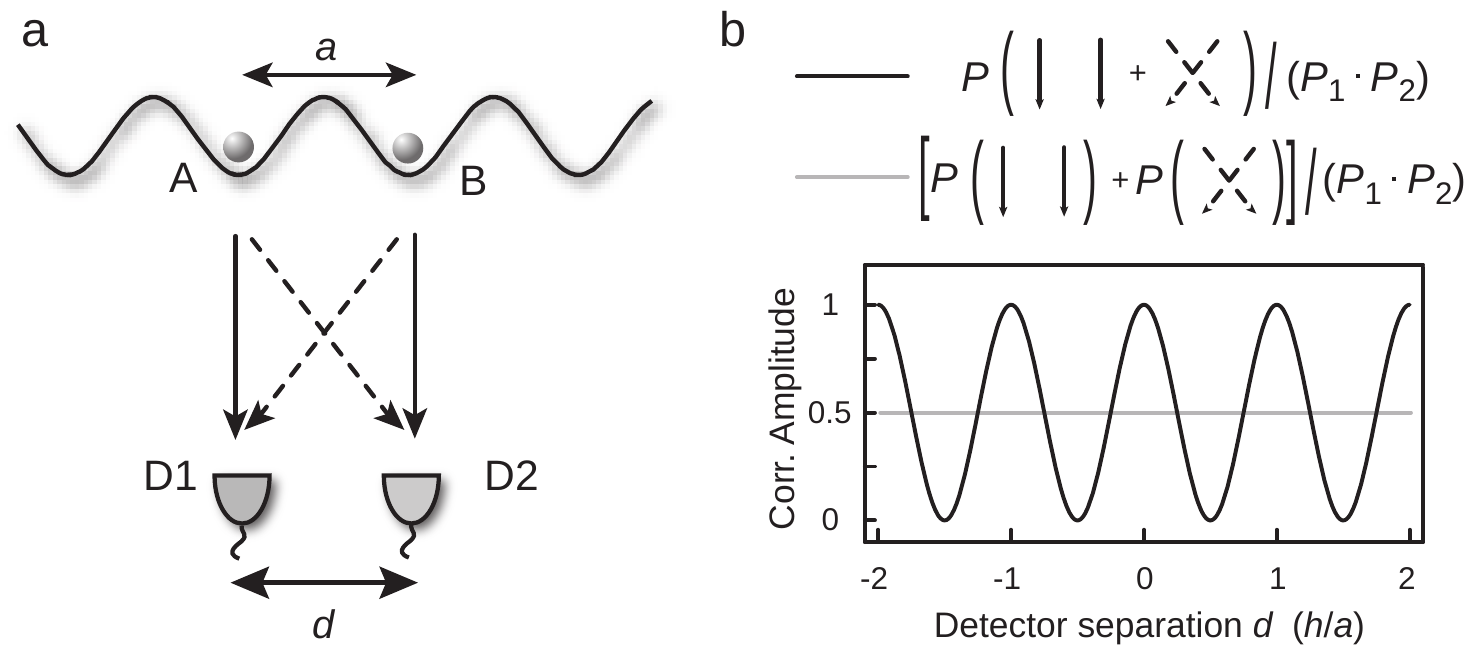}
	\caption{Origin of two-point correlations between the signals from two detectors at relative distance $d$. Atoms are emitted from two separate trap locations $A$ and $B$, and are detected by two detectors with respective probabilities $P_1$ and $P_2$. For indistinguishable particles, the amplitudes for the two possible scenarios for simultaneous detections have to be added, which have a distance-dependent relative phase. This leads to a modulation in the joint detection probability, resulting in a sinusoidal modulation of the correlation amplitude -- the joint detection probability normalized by the uncorrelated joint detection probability.}
	\label{fig:sf_CorrelationSchematics}
\end{figure}

Let us assume two atoms are prepared in two separate traps, which are indistinguishable, but strongly localized at two separate lattice locations, as illustrated in figure \ref{fig:sf_CorrelationSchematics}.
If these two locations are the states $\ket A$ and $\ket B$, then this corresponds to the state $\opbd_B \opbd_A \ket 0$. When these two particles are released from their respective traps, the corresponding wave packets will expand quickly, due to their initial strong localization. After an expansion time $t$ which is long enough that the particle from each source can reach each of two detectors $1$ and $2$ in locations $x_1$ and $x_2$, we use the detector signals and analyze the resulting statistics of the atom counts in the detector locations. Obviously, each detector can either detect two atoms, one atom, or none at all. If we assume that the probability of a single atom from either of the two sources to reach a detector in location $x$ is $P(x)$, then, for classical particles, one expects the probability for finding one particle in each of the two detectors to be $P(x_1)\cdot P(x_2)$. For indistinguishable particles however, this is not necessarily the case, as illustrated in figure \ref{fig:sf_CorrelationSchematics}. The fact that a particle was found in each of the two detectors is the result of a measurement, and thus the total quantum mechanical amplitudes of all possible paths which led to this result have to be added up prior to computing the probability by taking the absolute square of the total amplitude. As we will see, the result in this case is probability which is a sinusoidally modulated in the detector distance $d$.

This effect was first described by Hanbury Brown and Twiss\cite{hanburybrown56a,hanburybrown56b}, originally in a quite different context, namely that of classical electromagnetic waves. Hanbury Brown and Twiss wanted to measure the distance between independent astronomical radio sources, or the size of an extended source. In their approach, they assumed radio waves which are emitted independently from both sources, but detected simultaneously at the two detectors, and discovered the correlated behavior for such waves both in calculations and experiment. They subsequently also demonstrated the effect with optical photons, strongly increasing the interest in understanding the nature of coherent photon fields\cite{glauber63c}. In the following, we will consider the situation with many independent, incoherent sources of bosons, such as an optical lattice, and compute the correlation properties expected in that case. Density-density correlations can also arise in other scenarios and without lattices of course, and some of these cases will be discussed in section \ref{sec:sf_CorrelationExperiments}.

\subsection{Density--density correlations for released atom clouds\label{sec:sf_NoiseCorrelationsDerivation}}
The absorption imaging technique used in most experiments fundamentally measures densities of particles as a function of position in space. Our measurement is therefore described by the density operator $\opn(\vect x)=\opbd(\vect x)\opb(\vect x)$, evaluated for many locations at the same time. Here, $\opb(\vect x)$ is the bosonic particle annihilation operator at location $\vect x$. Of course the 3D distribution is projected to a 2D image, which we will have to consider later in order to predict the measured signals. For the time being, however, we neglect this projection, and assume that the measurement has the same dimensionality as the space in which the atoms expand, using vector notation for positions $\vect x$.

Evaluating the expectation value at time $t$ for the density operator for $N$ indistinguishable particles in different, but overlapping, modes indexed by $j$ and $k$, we obtain
\begin{equation}
\average{\opn(\vect x,t)}=\average{\opbd(\vect x,t)\opb(\vect x,t)}=\bigaverage{\sum_j\opbd_j(\vect x,t)\sum_k\opb_k(\vect x,t)}.
\end{equation}
Here, the position $\vect x$ corresponds to the location after time of flight. The modes $j$ are defined by their respective initial locations $\vect x_j$ at time $t=0$, from which they are released. Now we need to relate this operator to the state before the release. We can write in general
\begin{equation}
\opbd_l(\vect x, t)=\Psi(\vect x-\vect x_l,t)\opbd_l,
\end{equation}
where $\opbd_l$ is the creation operator of a particle at the initial state $l$. $\Psi(\vect x,t)$ is the wave function of a particle localized to the on-site wave function at time $t=0$ and freely expanding from there during the expansion time $t$. For the ground states of strongly confining potentials, such as deep optical lattices, the localized wave function on a site can typically be very well approximated by a Gaussian function (in the more general lattice case it would be the Wannier function)
$$\Psi(\vect x, t)=W(\vect x,t) \ee^{i\frac{\hbar t}{2 m^2 \sigma(0)^2} x^2/\sigma(t)^2}.$$
Here, $W(\vect x,t)$ is the 3D spherically symmetric Gaussian amplitude envelope with
$$W(\vect x, t)=\frac{1}{{(2\pi \sigma(t))}^{3/4}} \ee^{i\theta(t)}\ee^{-\frac{x^2}{2 \sigma(t)^2}}$$
which has a width $\sigma(t)$ evolving as
\begin{equation}
\sigma(t)=\sqrt{\sigma_0^2+\frac{\hbar^2 t^2}{\sigma_0^2 m^2}},
\end{equation}
and a slowly varying global phase $\theta(t)$ which in the computations in this chapter cancels and has no effect.

For long times $t$ much larger than the on-site oscillation period (given by the inverse band gap of the lattice), this simplifies to $\sigma(t)\approx \hbar t/\sigma_0 m$. This condition is typically extremely well fulfilled, so we have
$$\Psi(\vect x, t)=W(\vect x,t) \ee^{i\frac{m}{\hbar t} x^2}.$$

Using the $\Psi$ notation, we therefore obtain for the density operator
\begin{eqnarray}
\opn(\vect x,t)&=&\sum_{j,k} 
   \Psi^*(\vect x-\vect x_j) \Psi(\vect x-\vect x_k) \opbd_j\opb_k \nonumber \\
	&=&\sum_{j,k} 
   W^*(\vect x-\vect x_j) W(\vect x-\vect x_k) 
	\cdot\ee^{i\frac{m}{2\hbar t}(-(\vect x-\vect x_j)^2+(\vect x-\vect x_k)^2)} \opbd_j\opb_k.
\end{eqnarray}

The envelope function $W(\vect x)$ is smooth on the scale of inital atom distribution (for long times of flight), so $W(\vect x+\vect x_j)\approx W(\vect x)$ and we can simplify the expression to 
\begin{eqnarray}
\opn(\vect x,t)=
	\sum_{j,k} 
   |W(\vect x)|^2 
	 \ee^{i\frac{m}{2\hbar t}(-(\vect x-\vect x_j)^2+(\vect x-\vect x_k)^2 )} \opbd_j\opb_k. \label{eq:sf_DensityOperatorValue}
\end{eqnarray}

The density--density correlation operator as the product of the densities at two locations therefore leads to the slightly bulky expression
\begin{eqnarray}
&\opn&(\vect x_1,t)\opn(\vect x_2,t)= \nonumber \\
&=&\sum_{j,k,l,m} 
   |W(\vect x_1)|^2 |W(\vect x_2)|^2
	\cdot\ee^{i\frac{m}{2\hbar t}(2x_1(x_j-x_k)+x_k^2-x_j^2)+(2x_2(x_l-x_m)+x^2_m-x^2_l)}\opbd_j\opb_k\opbd_l\opb_m.\nonumber \\
	\,\label{eq:sf_SecondOrderCorrelationTOFGeneral}
\end{eqnarray}
The crucial part for the structure of the correlations is of course the operator product $\opbd_j\opb_k\opbd_l\opb_m$, and this is where the quantum properties of the particles enter. In order to evaluate the product, we first have to bring it into normal ordered form, which uses the commutation relation for bosonic operators $[\opb_j,\opbd_k]=\delta_{jk}$:
\begin{equation}\label{eq:sf_NormalOrdering}
\opb_j^\dagger \opb_k \opb_l^\dagger \opb_m = \hat b_j^\dagger(\hat b_l^\dagger \hat b_k + \delta_{lk}) \hat b_m = \opb_j^\dagger \opb_l^\dagger \opb_k  \opb_m + \delta_{lk} \opb_j^\dagger  \opb_m.
\end{equation}
The first term provides the second order correlator, whereas the second term corresponds to the correlation of each atom with itself, resulting in a strong peak for zero relative momentum (autocorrelation peak) which will be discussed later, but neglected here.

 For now, we will discuss only the case of indistinguishable bosons, but clearly this is where the difference between bosons and fermions enters: The fermionic anticommutation relation will lead to a minus sign before the second order correlation term, resulting in an inversion of the correlation signal, but not of the autocorrelation peak. The fact that the correlation features originate from the quantum commutation relation is the reason why this particular kind of correlation signal is called ``quantum noise correlations''. Other processes, such as collisions, can also lead to correlation features in the atom distribution without a direct quantum mechanical origin, as will be discussed later in this chapter.

\subsection{Correlations in particle ensembles from deep lattices\label{sec:sf_CorrelationFromLattice}}
We now have related the second order correlation operator after time of flight to the creation and destruction operators on the lattice sites. Now, we can evaluate this for a given state inside the lattice and determine the expectation value for the operator. We will mostly discuss the case of localized atoms with a fixed particle number $n_j$ on each site $j$ and undefined phase (Fock states), as this corresponds most closely to the interesting strongly correlated many-body states such as the Mott insulator. This is in contrast to he case of a BEC or superfluid phase inside the lattice, which has no defined particle number per site, but instead a well-defined phase.

Let us consider such a system of lattice sites where $n_j$ is well-defined on each site $j$ of the trap, with a size of $L$ lattice sites along a given direction, and $N=\sum_j n_j$. For a system consisting entirely of Fock states with particle numbers $n_j$, the operator $\opbd_j\opb_k=\delta_{jk} n_j$. Using this, we can evaluate the normal-ordered second order correlator as
\begin{equation}\label{eq:SFMottInsulatorCorrel}
 \langle\opb_j^\dagger \opb_l^\dagger \opb_k  \opb_m\rangle= \delta_{jm} \delta_{lk} n_j n_l +
\delta_{jk} \delta_{lm} n_j n_l + \delta_{jk} \delta_{jl} \delta_{jm} (n_j(n_j-1) - 2 n_j^2).
\end{equation}
The third term in this sum only adds an offset of the order $1/N$ to the final result and we will neglect it. The second term contributes to a Gaussian offset of the autocorrelation function and is canceled in the normalization, as we will see. The first term is the one which contains a non-trivial spatial structure, which we will determine in the following.

The contribution of the first and second term to the full expectation value is evaluated using equation \ref{eq:sf_SecondOrderCorrelationTOFGeneral}:
\begin{eqnarray}
&\langle&\opn(\vect x_1,t)\opn(\vect x_2,t)\rangle=\sum_{j,k,l,m} 
   |W(\vect x_1)|^2 |W(\vect x_2)|^2 \nonumber \\
	&\cdot&\ee^{i\frac{m}{2\hbar t}(2\vect x_1(\vect x_j-\vect x_k)+\vect x_k^2-\vect x_j^2)+(2\vect x_2(\vect x_l-\vect x_m)+\vect x^2_m-\vect x^2_l)}(\delta_{jm} \delta_{lk} n_j n_l + \delta_{jk} \delta_{lm} n_j n_l)\nonumber \\
&=&\sum_{j,l} 
   |W(\vect x-\vect d/2)|^2 |W(\vect x + \vect d/2)|^2
	\cdot(\ee^{i\frac{m}{2\hbar t}\vect d\cdot(\vect x_l-\vect x_j)}n_j n_l+N^2) \label{eq:sf_StructureTermSum}
\end{eqnarray}
In the last line, we introduced the relative distance of the two detectors $\vect d=\vect x_2-\vect x_1$ and their center of mass position $\vect x=(\vect x_1+\vect x_2)/2$. We can see that only the smooth envelope part of the function depends on $\vect x$, and the remainder only on $\vect d$. We therefore define our main correlation observable $C$ as the $\vect x$-integral of this expression, which then only depends on $\vect d$. In addition, we normalize the expression with the expectation value for uncorrelated particles, which is easily computed as the product of the density expectation values:
\begin{equation}
C(\vect d)= \frac{\int \langle \opn({\vect x}-\vect d/2) \opn({\vect x}+\vect d/2)\rangle d^3 x}
{\int \langle \opn({\vect x}-\vect d/2)\rangle \langle \opn({\vect x}+\vect d/2)\rangle d^3 x} -1.\label{eq:sf_NormalizedCorrelationIntegral}
\end{equation}
This expression can also be identified as the autocorrelation function of the density divided by the autocorrelation function of the expectation value of the density (and hence corresponding to an uncorrelated ensemble).

By inserting the two known contributions of the correlator, from equations \ref{eq:sf_StructureTermSum} and \ref{eq:sf_DensityOperatorValue}, we obtain as the expectation value for the localized atoms
\begin{eqnarray}
C(\vect d)&=&
\frac{\int 
|W({\vect{x}}-\vect{d}/2)|^2 |W({\vect{x}}+\vect{d}/2)|^2 \cdot (\sum_{j,l} \ee^{\frac{i m}{\hbar t} (\vect x_l-\vect x_j)\cdot\vect d} n_j n_l+N^2)\,d^3 x}
     {\int 
       |W({\vect{x}}-\vect{d}/2)|^2 |W({\vect{x}}+\vect{d}/2)|^2 N^2\,d^3 x}-1 \nonumber \\
     \nonumber \\
&=& \frac{1}{{N^2}} \sum_{j,l} \ee^{i\frac{m}{\hbar t} 
\vect d \cdot (\vect x_l-\vect x_j)} n_j n_l = \frac{1}{{N^2}} \left|\sum_{j} \ee^{i\frac{m}{\hbar t} 
\vect d \cdot \vect x_j} n_j \right|^2. \label{eq:sf_AutocorrelationExpectationvalue2}
\end{eqnarray}
In a lattice structure, the locations of the lattice sites $x_j$ are spaced at regular intervals. Therefore, this expression corresponds to the absolute square of a Fourier sum, which can be easily seen in one dimension if inserting $x_j=j\cdot a$:
$$C(d)= \frac{1}{{N^2}} \left|\sum_{j}  n_l \ee^{i\frac{a m j}{\hbar t} 
 d} \right|^2.$$
The periodicity of this sum is 
$$l=2\pi \cdot \frac{\hbar}{m a} t=2\frac{\hbar k}{m} t,$$
which is the distance corresponding to the reciprocal lattice momentum, or two times the lattice photon recoil momentum in the case of a typical retroreflected optical lattice as described in the previous chapters. The same holds in three dimensions, with three indices. 

\begin{figure}
	\centering
		\includegraphics[width=1.0\textwidth]{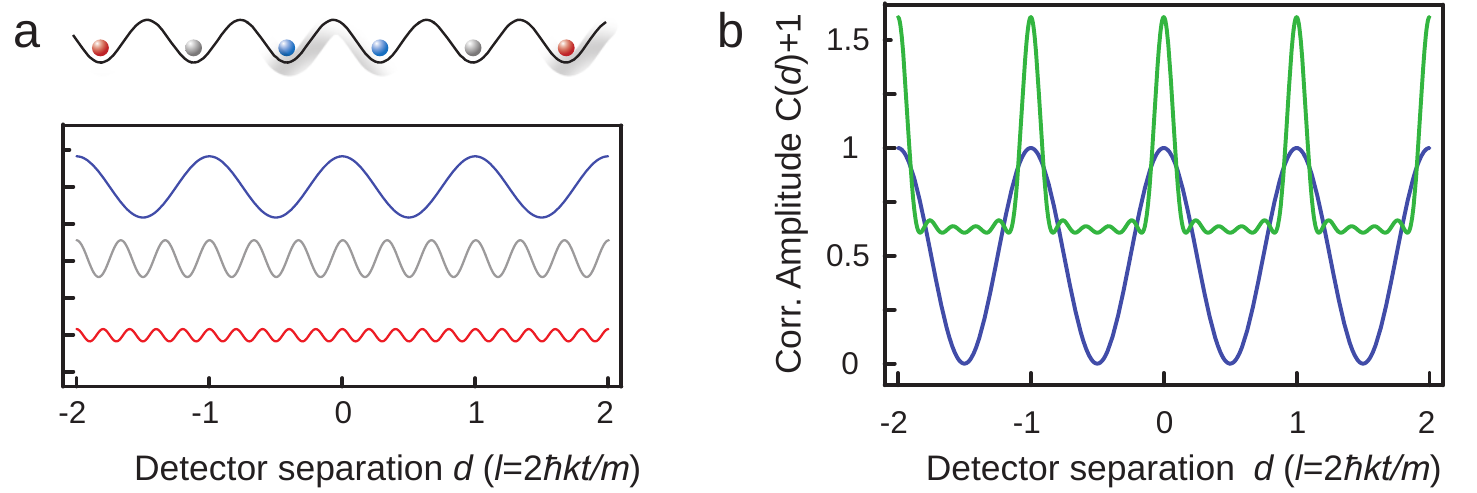}
	\caption{Correlation signals for pairs at various relative distances before release. In (a), pairs of atoms at a given distance are shown in different colors, which lead to a correlator expectation value each shown with the same color below. The different amplitudes of the components are caused by the fact that the number of pairs at a given distance goes down with distance for a finite chain. In (b), the sum of these components is shown, illustrating the appearance of a regular train of narrow peaks.}
	\label{fig:sf_Signal1D}
\end{figure}

Figure \ref{fig:sf_Signal1D} shows the relation of the atom distribution to the Fourier sum using an atom pair picture: A pair at a given distance produces a Fourier component at the inverse length, and the sum of the contributions from all possible pairs gives the total signal.

We can also directly evaluate the integral over one period of the complete function (in the general case in 3D)
\begin{equation}\label{eq:sf_SignalStrengthFromString}
S=\int C(\vect d) d^3d=\frac{l^3}{N^2}\sum_j n^2_j,
\end{equation}
because the integral over a squared Fourier sum is the sum of the squared Fourier coefficients. For a constant on-site density, we can therefore see that the signal will scale with the atom number as $1/N$. This can be expected from the fact that the signal originates from the shot noise fluctuations of particles hitting the detector: This signal scales as $\sqrt N$, the second order correlation therefore as $N$ - but the normalization is a product of densities and thus scales as $N^2$. The same normalization of course also means that the random noise in the correlation signal will also scale as $1/N$, so the signal to noise ratio is not decreased, as long as atom shot noise is the dominant contribution.
For the case of a homogeneous distribution of one atom per site and therefore $L=N$ sites in 1D, the Fourier sum results in the correlation pattern
\begin{equation}
C_{1D}(d)=\left|\sum_j\frac{1}{N} \ee^{i\frac{2\pi}{j}d}\right|^2=\frac{\sin(\pi N d/l)^2}{\sin(\pi d/l)^2}
\end{equation}
which is shown in figure \ref{fig:sf_Signal1D}(b) for the case of two and 6 sites. A series of sharp peaks appears, spaced by a distance $l$, with a height of 1 and a width which decreases as $1/N$ with increasing number of sites.

As can be seen from this result, there is no envelope to this function due to the normalization, therefore the range over which momenta can be correlated will be limited by the increasing noise outside of the original envelope function.

In terms of the resolution of the method -- the minimum size of a given feature in the signal -- there is an important difference to the imaging of density-related quantities: In contrast to the usual case of determining density profiles, the far-field approximation has only been made for the Wannier envelope part of the wave functions, not for the phase term. As a consequence of this, the relations derived here are typically valid before the ensemble has expanded into the far field. A specific consequence of this is that the correlation peaks for example can be smaller than the initial size of the cloud: In contrast to what is described in chapter \ref{sec:imagingYYZ} for  typical, direct momentum distribution measurements, the signal obtained is not effectively convolved with the initial ensemble size prior to release.

\section{Noise correlations and actual experimental implementation}
In the previous part of this chapter, we have seen the quantum mechanical origin and derived the shape of the correlations between the densities of different locations after the cloud has expanded from the trap. This has been done for the theoretical, ``perfect'' density observable, which has no additional noise, and perfect resolution. This is of course not the case in real experiments, so in order to derive the properties of a correlation signal which can be measured in an actual experiment, we have to take at least the intrinsic properties of actual implementations of the scheme into account. We will then describe how the signal is extracted from the data which an experiment produces, and discuss the main features of the signal which is obtained.

\subsection{Noise correlations and optical detection}

The measured atom distributions after expansion in the usual case are measured by optical absorption imaging. The cloud is illuminated from one side with a resonant illumination beam and the corresponding absorption profile is projected to a camera. From the camera data one obtains a 2D array of optical densities, as a function of position. In this way, the column density of atoms can be computed as outlined in chapter \ref{sec:imagingYYZ}. The two-point correlation will then be determined by correlating the densities measured between two points in this detector plane. 

\begin{figure}
	\centering
		\includegraphics[width=1.0\textwidth]{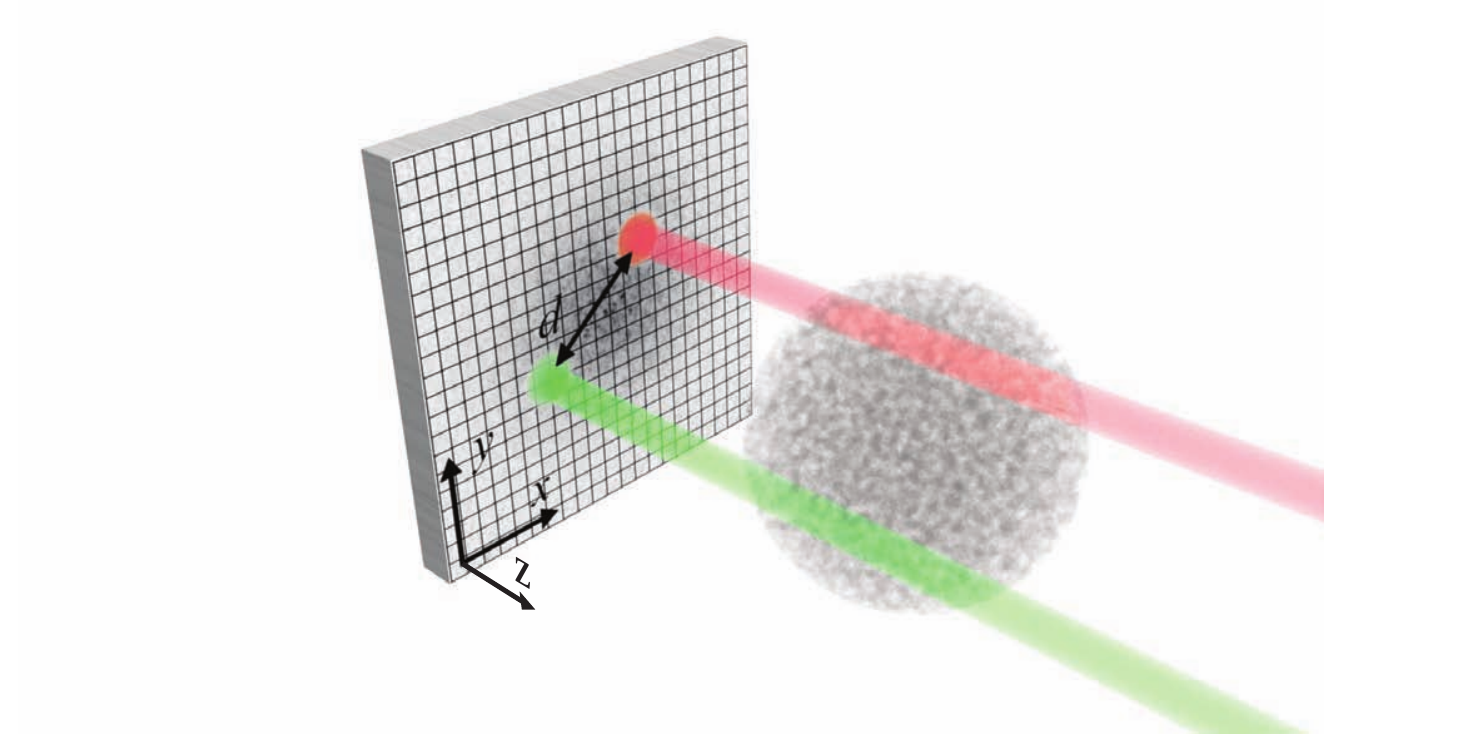}
	\caption{Schematic view of the optical detection of correlations. The shadow of the atom cloud is projected onto a CCD array. The density of the atoms is integrated by the imaging system within the ``bins'' along the $z$ axis whose size is determined by the resolution of the imaging system.}
	\label{fig:sf_CCDDetectorSchematics}
\end{figure}

This scheme is illustrated in figure \ref{fig:sf_CCDDetectorSchematics} where the volume of the two columns which are effectively being correlated are shown in green and red.
This also illustrates three fundamental limitations of such an imaging method: Firstly, the detector is two-dimensional and the absorption process therefore always integrates the density along the direction of propagation of the imaging light. Secondly, the picture elements, the pixels, have a finite size, so in any case the signal is in addition integrated over the surface area of a pixel. The detection volume is therefore finite, where before we have always discussed detection at specific points.

Thirdly, as illustrated in the figure, usually the resolution of the imaging system is such that the smallest possible feature which can be imaged is actually larger than the pixel size. This is by design, as one effectively chooses pixel sizes small enough as to be limited only by the optics, not the pixel area. The shape and size of the integration area that is effectively realized by the imaging system is defined by the point spread function (PSF). It is defined as the detector output for the case of a perfect point source in the object plane of the imaging system. The pixelized 2D density distribution is effectively convolved with the PSF of the imaging system to form the measured image. It is this point spread function which therefore determines the volume in the detection region, within which all atoms are integrated. The measured shot noise is then the noise of this integrated value, typically corresponding to several pixels (which explains why the shot noise seen in figure \ref{fig:sf_GaussianImage} is much lower than that expected for a column density of 30 atoms per pixel). 

Of course, the detection regions corresponding to adjacent pixels will overlap in such a case. An atom will therefore contribute signal to several pixels of the image -- all those within the range of the PSF -- leading to a correlation of the signal in adjacent pixels even without correlations between atoms. Effectively, we can therefore take the absorption and the point spread function into account by a convolution of the density distribution with the PSF in the $x$ and $y$ directions, and integrating over the entire cloud along the $z$ direction:
\begin{eqnarray}
c(r_x,r_y)&=&\int n(r_x+x',r_y+y',z) PSF(x', y') dx' dy' dz\nonumber \\
&=&\int n(\vect r+\vect x') PSF(\vect x') d^3 x'
\end{eqnarray}
The measured column densities now correspond to the values of this quantity at the center of each pixel. In the second line the expression was simply rewritten in vector form, with $r_x$ and $r_y$ being the $x$ and $y$ components of $\vect r$.

For the correlation signal defined in equation \ref{eq:sf_NormalizedCorrelationIntegral}, we have to do the same transformation (here, $\otimes$ denotes the convolution operator):
\begin{eqnarray}
C(d_x,d_y)&=&
\frac{\int \int \int \langle PSF(\vect x'-{\vect x}+\vect d/2)n(\vect x') PSF(\vect x''-{\vect x}-\vect d) n(\vect x'')\rangle d^3 x' d^3 x'' d^3{x}}
     {\int \int \langle PSF(\vect x'-{\vect x}+\vect d/2) n(\vect x') d^3 x'
      \rangle \int \langle PSF(\vect x''-{\vect x}-\vect d) n(\vect x'') \rangle d^3 x'' d^3{x}}-1 \nonumber \\
      &\approx&
      \int \int PSF(\vect x'-\vect d)PSF(\vect x''-\vect x')C(\vect x'') d^3 x'' d^3 x' \nonumber \\
      &=&(PSF \otimes PSF \otimes C)(\vect d)
\end{eqnarray}
To predict the expected correlation signal from such a detector, we therefore need to convolve the correlation expectation value for point detection twice with the imaging system point spread function. The $z$ integration is included in $PSF(x,y,z)$ in the sense that the function does not depend on $z$. As the relevant part of the correlation signal is periodic, the $z$-integration can be effectively limited to the integration over one period.

Due to this convolution with the point spread function, a meaningful measure of the strength of the correlation signal is therefore the integral of the signal under the peak. This value is unchanged by convolution, and will therefore correspond to the integral of the theoretically predicted peak, independent of imaging resolution.

\begin{figure}
	\centering
		\includegraphics[width=1.0\textwidth]{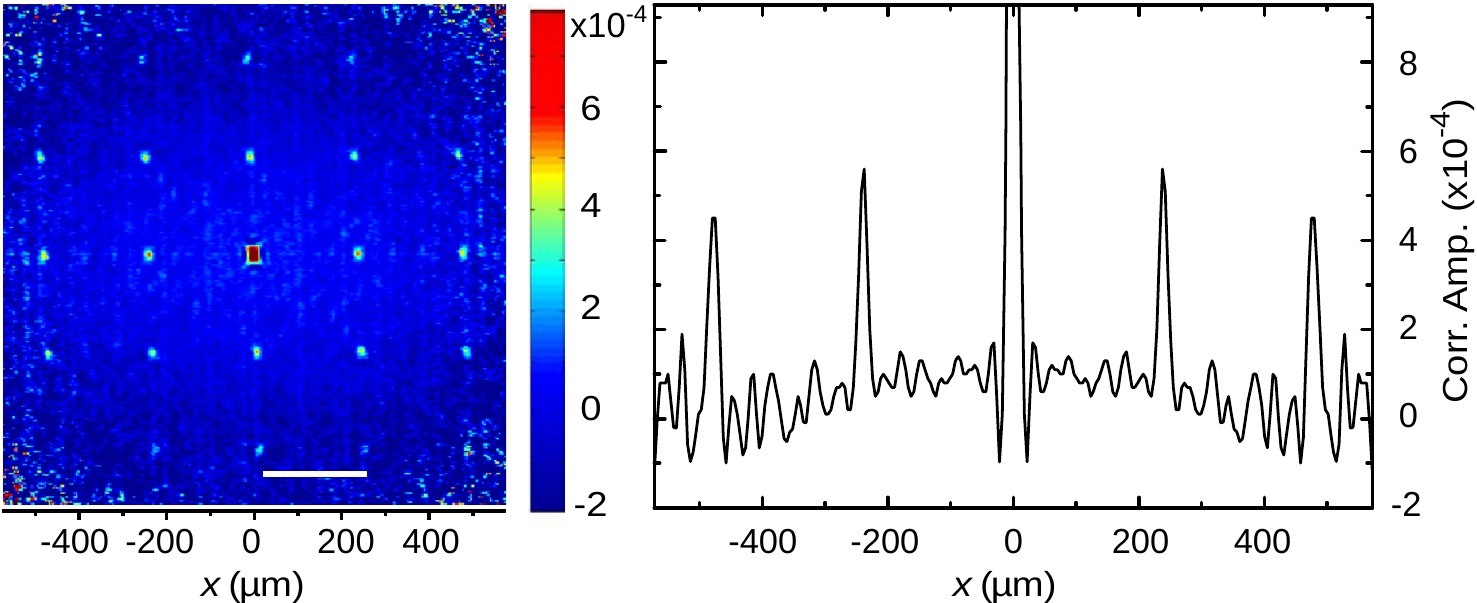}
	\caption{2D noise correlation signal $C(\vect d)=C(x,y,z)$ from a 3D deep optical lattice, with a horizontal profile through the central line of the 2D correlation function. The central autocorrelation peak is clearly visible, the correlation signal has an amplitude of approximately $4\times10^{-4}$. (figure adapted from \cite{foelling05}, Nature
434, 481, (2005))}
	\label{fig:sf_CorrelationImage}
\end{figure}

An actually measured correlation function for a Mott insulator in a deep optical lattice\cite{foelling05} is shown in figure \ref{fig:sf_CorrelationImage}. One can see that the measured amplitude is much smaller than 1, mostly due to the $z$ integration because the Mott insulator consisted of many planes of atoms along the $z$ direction. The imaging system's point spread function in the lateral direction is approximately twice the effective pixel size.

\subsection{Structure and strength of noise correlation signal\label{sec:sf_CorrelAmplitudeDiscussion}}
Let us discuss the signal obtained in a measurement as shown in figure \ref{fig:sf_CorrelationImage}. The expected sharp peaks derived in the previous section show up at the expected locations given by integer multiples of the lattice momentum range, $2 \hbar k_{lat}$, where $k_{lat}$ corresponds to the lattice light $k$.
Theoretically, the function $C(\vect d)$ has a constant background of 0, but the denominator will decrease for large values of $\vect d$, the `field of view'. Therefore, the range of momenta over which correlations can be detected is effectively limited by the increasing noise toward the edges of the data, as can be seen in the corners of figure \ref{fig:sf_CorrelationImage}.

The strongest feature in the signal is the self-correlation peak in the center of the image. It is several orders of magnitude larger than the actual quantum noise correlation, as every atom is perfectly correlated with itself. However, its shape is determined by a convolution of the point spread function with itself, which allows for independent determination of the PSF shape.

The actual, regularly spaced correlation peaks have an amplitude of approximately $4\times10^{-4}$. As discussed before, the expected signal in 3D prior to convolution and column integration for unity filling in the trap is $1$, with a width $w$ such that the total weight of the signal is $S$. The $z$ integration, effectively averaging over one period of the signal, reduces the weight of the peak to $S\cdot w_z/l$, where $w_z/l=L_z$ corresponds to the number of planes $L_z$ occupied along the $z$ direction. If the PSF is wider than $w$, and therefore determines the measured width of the correlation peaks, the convolution and $z$ integration will thus lead to a constant width signal with further decreased amplitude such that the weight of the peak is $S_{xy}$. 

The expected magnitude for $S_{xy}$ for a system with constant initial filling $n$, size $L_{x,y,z}$ in the $x$, $y$ and $z$ direction respectively, is therefore
\begin{equation}
S_{xy}=\frac{l^2}{N \cdot L_z}n^2. \label{eq:sf_CorrelPeakWeight2D}
\end{equation}
As $l$ scales linearly with the expansion time, the signal scales quadratically in $t$, and as $1/N$ with atom number $N$. Larger filling factors lead to larger signals at constant atom number.

The $1/N$ scale of the correlation amplitude can also be shown with another, more intuitive interpretation: As the normalized correlation signal is derived from the shot noise on each detector bin with a second order function, it will scale with this shot noise squared. The normalized shot noise is given by $1/\sqrt{N_{bin}}$, with $N_{bin}$ the number of particles in the corresponding detection bin. The expected amplitude therefore must scale as $1/N_{bin}$ for a constant density source.

To obtain unity signal strength therefore requires a reduction of the atom number to the order of less than one atom per bin, and the reduction to a single plane of lattice sites, to avoid the signal reduction from $z$ axis integration. This has been realized\cite{simon2011a} for a 1D system by implementing the expansion of a chain of atoms along one axis. The resolution of the imaging system here is such that after expansion, on average much less than one atom occupies a detector bin. In this case, a modulation depth of the correlation signal from a Mott insulator of around 0.3 was obtained, most likely limited only by the inhomogeneity of the potentials and by interactions of the atoms during expansion along the tightly confining tubes. This experiment will discussed in some more detail in section \ref{sec:sf_PatternedLoading}.

For small in-trap ensembles, with a low number sites along a given direction contributing to the correlation signal, long expansion time and good imaging, the lateral size of the correlation peak can also be resolved for 3D time of flight experiments with the standard absorption imaging technique. This has been used to characterize the interacting Bosons in a 2D lattice system through the superfluid to Mott insulator transition\cite{spielman07}. During the transition, the formation of the insulating part of the ensemble was tracked by analyzing the amplitude and size of the correlation signal. A difficulty with this kind of measurement is the fact that a partially coherent system will exhibit interference peaks already in the expectation value of the density, with the same periodicity as the second order correlation signal. Small fluctuations in the strength and position of these interference patterns will then generate second order noise correlation signals which are not removed by normalization. In order to avoid this, the parts of the images affected by the diffraction pattern can be excluded from the correlation analysis \cite{rom06,spielman07}, allowing for the analysis of both second order noise correlations and the coherence pattern in the density expectation value, from the same images.

\subsection{Experimental implementation and data analysis}

As most experiments are by default equipped for time of flight absorption imaging, the use of noise correlation analysis for optical lattice systems as introduced in the previous chapter is typically quite straightforward.
In terms of hardware and even experimental sequences, no fundamental changes have to be 
made to typical setups.  

As outlined in chapter \ref{sec:sequenceYYZ} a typical experiment will start with a preparation phase, during which the atoms are cooled to the required low temperature and low entropy.
This cold ensemble is subsequently loaded into a trap with the desired configuration, which in itself often constitutes the experiment. For more complex sequences, additional manipulation steps or periods of in-trap time evolution might follow. After this, the time of flight expansion follows, with absorption imaging to determine the momentum distribution of the resulting ensemble, and the cycle is started over.

For implementing noise correlation analysis on such a system, none of these steps have to be modified in principle, it is essentially just a different method of analyzing the data from such experiments. This data is a set of absorption images, which each correspond to the 2D column density distributions of a single realization of the expanded atom cloud.
The main feature of noise correlation analysis is that it focuses on fluctuations of the data rather than the expectation values. As a consequence, while combining data from several (typically many) experimental runs for identical parameters is required for identifying the small amplitudes of the correlations within the intrinsic noise, this data can not just be averaged and then analyzed, as is often done with expansion images to improve the signal to noise ratio. Instead, correlations within the fluctuations are computed, first by applying expression \ref{eq:sf_NormalizedCorrelationIntegral} to the data obtained from the experiment. This is fairly straightforward: The pixel-pixel correlation function defined as the autocorrelation integral in 2D
$$ A(\vect d)= \int n({\vect x}-\vect d/2) n({\vect x}+\vect d/2) d^2{x}$$
can be directly evaluated as a sum over the pixels of the image
$$A(\vect d)=\sum_{\vect x_1,\vect x_2; \vect x_1-\vect x_2=\vect d} n(\vect x_1)\cdot n(\vect x_2),$$
where $\vect x_j$ are now discrete 2D vectors in the imaging plane, and $n(\vect x)$ is the measured column density at the pixel location $\vect x$.

For the normalization, the autocorrelation function of the expectation value has to be determined. For this, the average over all images from the dataset is usually used as an approximation for the expectation value, of which the autocorrelation function is then computed. After normalization, the correlation function can be analyzed for example by fitting the correlation peaks' sizes and weights.

\subsection{Numerical considerations}
Typically, several dozens of images are necessary to average the atom and photon shot noise to the point where the signal to noise ratio is acceptable. This can make the calculation of the correlation functions computationally quite heavy. A standard technique to compute the autocorrelation function of the cloud is therefore to apply the Wiener-Khintchin theorem which allows for the direct autocorrelation (requiring on the order of $\frac{1}{8} N_\mathrm{p}^2$ multiplications, where $N_\mathrm{p}$ is the number of pixels in the data) to two 2D Fourier transforms, which can scale as $N_\mathrm{p}\cdot \log(N_\mathrm{p})$ instead. Some of this gain is lost, as the Fourier--type algorithm fundamentally ``wraps around'' the data at the edges (cyclic convolution), generating invalid terms in the autocorrelation sum due. To avoid this, the data has to be padded with zeros along both axes, quadrupling the number of pixels (if the entire bitmap is being analyzed). Nonetheless, typically implementations using fast Fourier transforming techniques will be much faster than direct multiplication due to the much better scaling.

\subsection{Alternative: correlated 3D single particle detection\label{sec:sf_3DMetastableCorrelation}}
So far, in the entire discussion on noise correlations we have considered the case where the atom density distribution is recorded as a 2D ``image'' which integrates out one direction of space, resulting in column densities. Additionally, normally many atoms reside in each effective detector volume, which itself is typically wider than a pixel size.

\begin{figure}
	\centering
		\includegraphics[width=1.0\textwidth]{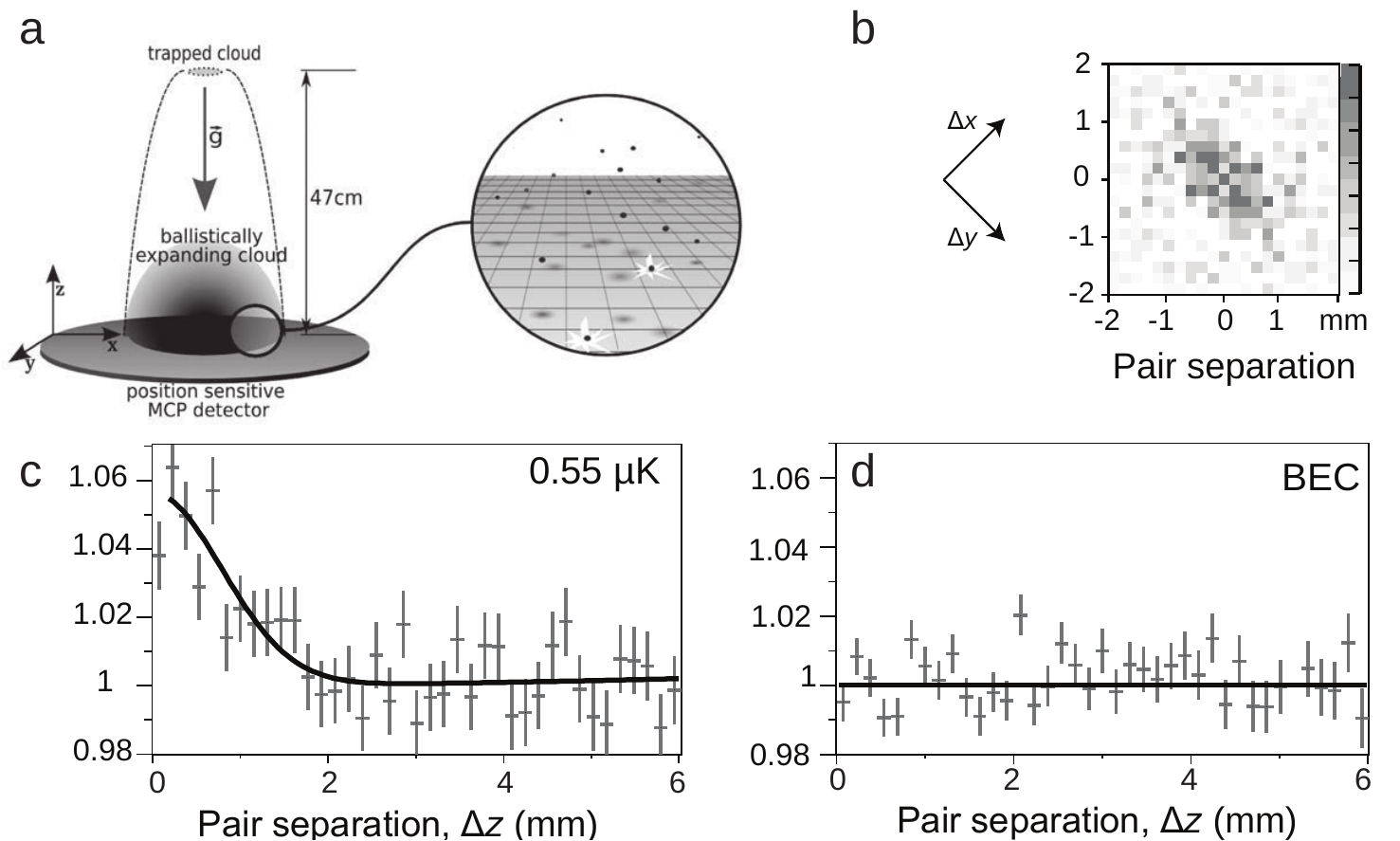}
	\caption{(a) Schematic drawing of a metastable Helium 3D detection setup. The atoms are released from the trap and expand ballisitically, until the cloud drops onto a multichannel plate (MCP) after about 320\,ms. The MCP detects particles impacting the surface with both spatial and temporal resolution, effectively providing 3D detection. A 2D slice of the correlation signal for a thermal gas of bosons is depicted in (b), which shows the bosonic bunching as a positive correlation peak for small pair separations. Because of the direct particle detection, there is no autocorrelation peak for zero distance, therefore the disappearance of the HBT correlation peak on the transition from a cold classical gas of bosons (c) to the flat, uncorrelated signal of a BEC (d) can be observed. (Figure adapted from \cite{schellekens05a}, with permission)}
	\label{fig:sf_HeliumCorrelations}
\end{figure}

An interesting scenario is one where the integration over a finite volume does not happen. Instead, the atoms could be detected individually after time of flight rather than being integrated into a density distribution. Indeed this is possible, and ideally even with full 3D information (but typically at the expense of reduced resolution compared to optical methods). Such systems have been implemented using ultracold metastable helium  (He$^*$) ensembles\cite{schellekens05a,gomes07a}, with the general concept illustrated in figure \ref{fig:sf_HeliumCorrelations}(a). For a more detailed description of He$^*$ experiments see ref\cite{Vassen2012}. Due to the fact that these atoms can be prepared in a metastable state with 20\,eV of internal excitation energy, they can be detected electrically rather than optically: the expanding atom cloud falls onto a so-called ``multi-channel plate'' (MCP), where the energy release from the metastable state upon contact with the detector causes one ionization event for each detected atom. The resulting electric charges are then amplified and detected electronically. These detectors can be spatially and temporally resolving using a delay-line technique. In such a case, rather than returning an average density per location, the detection system generates a ``list'' of all detected atoms including  detection time and 2D location per event. It is instructive to consider the difference this will make to the correlation analysis.

One important difference is the fact that, making the same assumptions about long time of flight as before, the $z$ component of the momentum can be reconstructed from the arrival time of the atom at the detector. This means that no integration along a column has to happen, the full 3D correlation function can be obtained.

More importantly yet, direct access to the individual particles is obtained - now we can operate on particles rather than densities. Therefore, no density noise is analyzed in this case (so technically this is not a ``noise correlation'' measurement in the density noise sense), the ``noise'' between the individual experimental runs appears in the sense that a completely different set of random atom positions is obtained on each realization.

Now, one can very easily use the same kind of analysis as discussed before, by just integrating all the different events of a single shot into a 3D (or even 2D) density distribution. The resulting density distribution can then be further processed as outlined above, except that full 3D information can be used. However, instead, it is also possible to correlate the recorded events directly, by counting the number of event pairs at a given distance $\vect d$. The direct identification of pairs makes for an important difference, as the correlation operator probed is different: Rather than measuring a correlation between densities $\average{\opn(x_1)\opn(x_2)}$, the counting of event pairs corresponds directly to the normal-ordered second order correlation function $g^{(2)}(x_1,x_2)$ form $\average{\opb^+(x_1)\opb^+(x_2)\opb(x_1)\opb(x_2)}$. Therefore, the autocorrelation peak, which resulted from normal ordering the operators in the density-density correlator, will not appear in such an analysis. As a practical consequence, this allows for measurements of correlations even for very small relative momenta, which could otherwise be obscured by the much stronger autocorrelation peak. For this reason, experiments using this technique have been able to explore direct (local) bunching and anti-bunching ($\Delta x \approx 0$) of bosons (Shown in figure \ref{fig:sf_HeliumCorrelations}b-d) and Fermions in bulk systems, without optical lattices, and even for higher order correlations than second order\cite{Hodgman2011}. 

\section{Experimental influences on the correlation signal}
Apart from the fluctuations caused by the shot noise, correlated as well as uncorrelated, fluctuations of any other type influence the detected correlation signal. In the following we will discuss several important effects that originate from the actual implementation and can affect the signal or even completely mask it.

\subsection{Atom number fluctuations}
In a typical experiment, the atom number will vary from one realization of the cold ensemble to the next. Such variations can be caused by any number of sources in the preparation procedure of the quantum gas, and can range from the 1\% level to arbitrarily high values. As the noise analysis method effectively classifies any deviation of the measurement from the expectation value as ``signal'', these atom number fluctuations influence the computed correlation function. In this case, however, the influence is not dramatic: a globally fluctuating density will simply result in a constant (positive, because the fluctuation has the same sign everywhere for a given image) offset of the correlation result, at least for small fluctuations. It can therefore be simply subtracted in the end, but alternatively, all images can also be normalized to the average atom number of the dataset in order to avoid the effect in the first place. For large fluctuations, one needs to consider that the signal scales non-linearly with the atom number as will be discussed in section \ref{sec:sf_inHomDensityDiscussion}, therefore the average signal does not necessarily correspond to the average atom number. Of course, filtering the dataset to a subset of the images with selected range of atom numbers can always reduce fluctuations at the expense of longer measurement times.

\subsection{Lattice geometry fluctuations}
\begin{figure}
	\centering
		\includegraphics[width=1.0\textwidth]{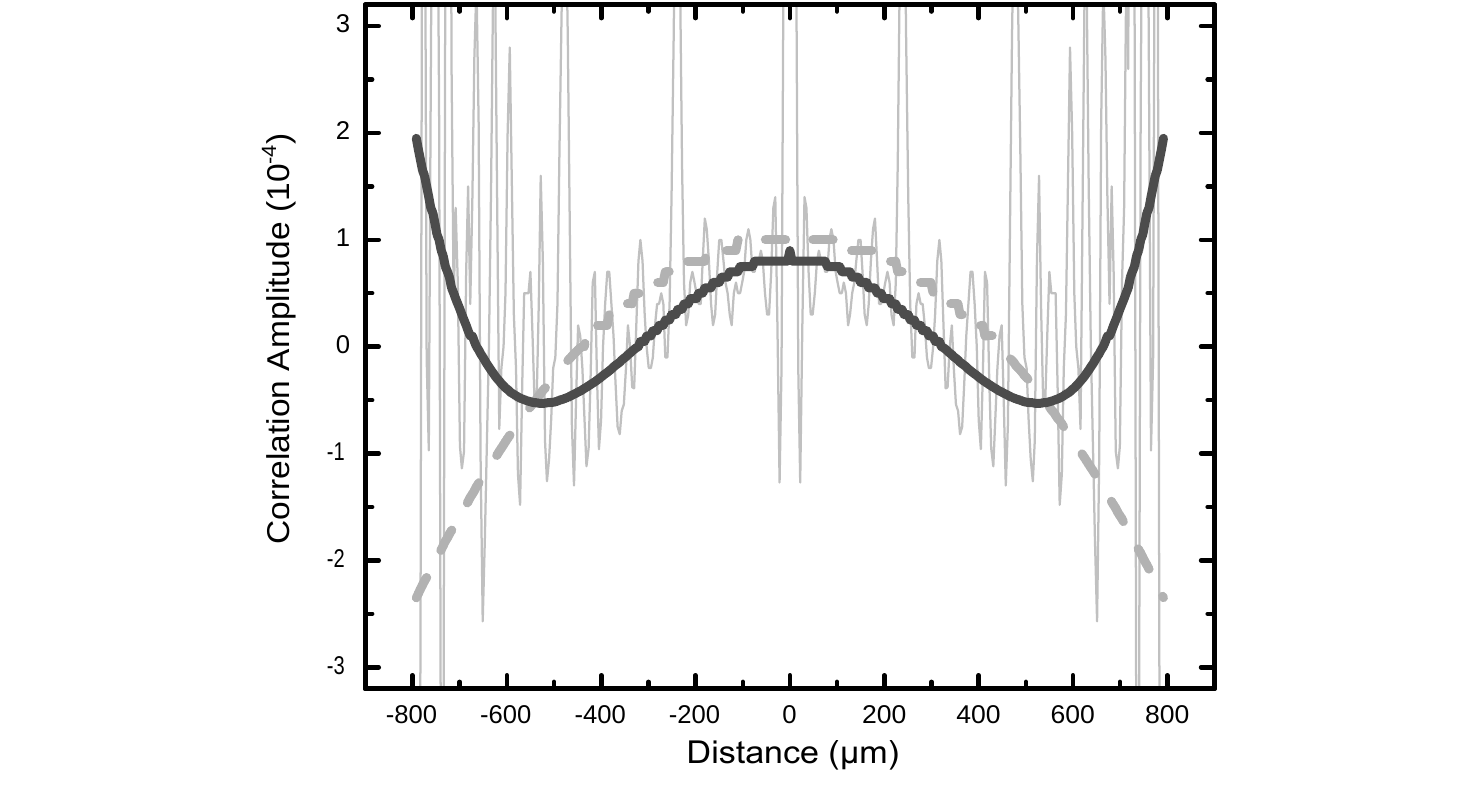}
	\caption{Influence of geometry fluctuations (position and envelope size shown as dark grey dashed and solid lines, respectively) of the expanded atom cloud. The RMS of the position variations is 2.2\,$\mu$m, the RMS of the cloud size fluctuation is 0.5\%. The light grey line shows data from an experiment, to which these values were approximately fitted.}
	\label{fig:sf_CorrelCloudFluctuations}
\end{figure}

In a similar way as the atom number, fluctuations of the trap configuration will cause correlated density variations which will be picked up by the correlation analysis process. These effects include fluctuations in the position of the atom source (and therefore of the entire expanded cloud) as well as fluctuations of the size of the cloud. Of course, movements of the camera position or a fluctuation in the expansion time will also lead to position and size variations. It is easy to see, that, for a moving density pattern, all points at which the derivative of the density along the direction of motion has the same sign, will be correlated; any pair of points with opposite sign in the derivative will be anticorrelated.
For a smooth atomic cloud shape, such variations are correlated on a large scale corresponding to the size of the system: A moving Gaussian cloud has a negative correlation between the two slopes of the Gaussian along the direction of the motion. Similarly, if the envelope function size fluctuates, there is a positive correlation between the two sides of the envelope. Figure \ref{fig:sf_CorrelCloudFluctuations} shows the resulting signal due to such fluctuations on a typical technical scale. For lattice correlations with a long correlation length these can be easily separated from the actual signal because much lower spatial frequencies are involved, but for smaller samples and short correlation lengths this effect can become more relevant.

\subsection{Detector imperfections and technical noise}
On the technical side, despite the fact that the signals detected can be in the $10^{-4}$ range or lower, the requirements on the detection system technology are not very stringent. For typical absorption imagery using scientific CCD sensors, the atomic and photon shot noise dominate the image noise. The averaging process needs to be long enough to average this noise below the signal level, so any additional, spatially uncorrelated, technical noise below this threshold will typically be averaged alongside these and show up in the central pixel of the correlation function. This, however, does not apply to all those noise components with spatial correlations. These can be caused by various effects in the imaging system and even by the camera itself. A typical example is a periodic distortion on the electronic readout channel of the sensor, for example due to crosstalk from a high frequency signal. Due to the sequential nature of the digitization of typical CCD camera systems, this translates to a ``brightness'' modulation which is periodic in the readout direction and can have large correlation lengths. In the normalized correlation data, this will then show up as a periodic background signal. The signal amplitude is that of the original fluctuation relative to the average signal - therefore an amplitude of less than one bit of camera signal can be detectable. For this reason, it is crucial that the CCD system does not suffer from correlated electronic noise, and it needs to be protected from electronic interference e.g. from RF sources.

Another typical scenario is interference fringes in the imaging system. Such fringes are very common in coherent illumination, caused for example by optical interference with reflections from vacuum chamber viewports. In typical absorption imaging setups using reference images, most of these fringes are suppressed by the image normalization -- the division of signal and reference images to obtain the column density as shown in expression \ref{eq:TorontoOD} in chapter \ref{sec:imagingYYZ}. However, small fluctuations in the interference patterns such as phase shifts cause an imperfect suppression, resulting in a weak fluctuating pattern which will be picked up by the noise analysis, as it has long correlation lengths. Typically, this effect will appear as a periodic structure in the correlation result, which can have very large amplitudes compared to the shot noise part of the signal. Such interferences must therefore be suppressed as much as possible.

\subsection{Interactions and inhomogeneous in-trap density\label{sec:sf_inHomDensityDiscussion}}

Using the theoretical expression for the density-density correlator, the expected amplitude for the correlation signal can be easily computed. As shown in section \ref{sec:sf_CorrelationFromLattice}, the normalized correlation signal actually reduces for increasing atom number $N$ as $1/N$ for constant density. However, in a trapped ensemble, the density is affected by the trapping potential and atom-atom interactions, the quantum statistics of the particles, and the temperature. From expression \ref{eq:sf_CorrelPeakWeight2D}, we can see that the normalized correlation signal increases with the on-site density, for constant total particle number. In the bosonic case discussed there, where multiple occupancies of lattice sites are not suppressed, the signal is therefore higher if the atoms are compressed to fewer sites.

\begin{figure}
	\centering
		\includegraphics[width=0.8\textwidth]{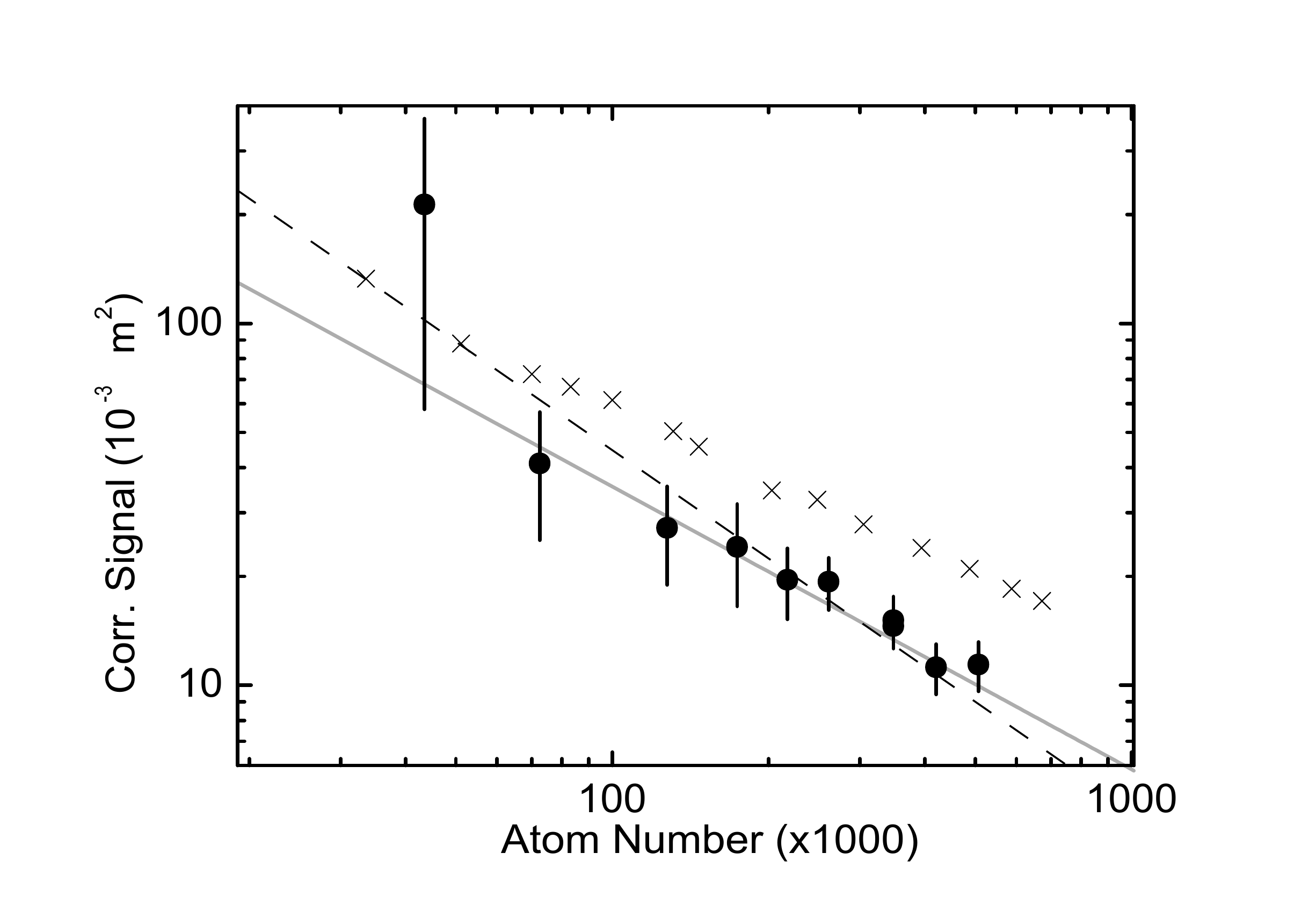}
	\caption{Strength of expected and measured correlation signal for a Rubidium Mott insulator after 22\,ms time of flight expansion. Black dots denote measured area under peak ($S_{xy}$), gray line is a power law fit to the data, dashed line is the calculated signal for a Mott insulator with given atom number and unity filling, and black x symbols denote the calculated signal for a Mott insulator with the expected shell structure for the real trapping potential employed. (figure adapted from F{\"o}lling et al \cite{foelling05}, Nature
434, 481, (2005)) }
	\label{fig:sf_SigVsAtomNumber}
\end{figure}

In figure \ref{fig:sf_SigVsAtomNumber}, this is illustrated by contrasting the expected unity-filling signal of a 3D Mott insulator at zero temperature with the expected signal from a Mott insulator in a realistic trap shape with increasing density in the trap center as a function of total atom number $N$. As additional Mott shells (see chapters \ref{chapterKollath} and \ref{sec:Weitenberg_ShellStructure}) form and the expected filling deviates from unity filling\cite{jaksch98}, the computed correlation signal deviates from the $1/N$ unity filling theory. The resulting modified slope is approximately reproduced by the measured values, which follows a fitted $N^{-0.78}$ power law for large atom numbers. However, all measured values are approximately half that expected from theory. This reduction is typical and has also been seen in other bosonic noise correlation measurements with large $N$. 

It is clear that, for interacting bosons, some reduction is expected from collisions between atoms during the expansion process. Such collisions remove the atoms involved from contributing to the correlation signal, as their momenta are effectively randomized unless the momentum transfer during the collision was smaller than the resolution of the detection scheme. Upon normalization of the signal with the entire population of atoms, including those which collided, the signal is therefore reduced by the corresponding factor. This interpretation gains further weight by the fact that experiments which used noninteracting polarized fermions gave values consistent with the full expected signal \cite{rom06}.

\section{Experiments employing noise correlation methods\label{sec:sf_CorrelationExperiments}}

\subsection{Bosons vs. fermions}
As discussed in section \ref{sec:sf_NoiseCorrelationsDerivation}, the difference between bosons and fermions is specifically visible in the two-particle correlator term. For the noise correlation terms discussed, the sign of the commutator appears in the correlation expectation value directly, inverting the structure of the correlation signal when switching from bosons to fermions. This has been shown by loading a sympathetically cooled, polarized Fermi gas into an optical lattice otherwise comparable to the one used with bosons, and subsequently measuring the noise correlations in expansion\cite{rom06}. 
\begin{figure}
	\centering
		\includegraphics[width=0.7\textwidth]{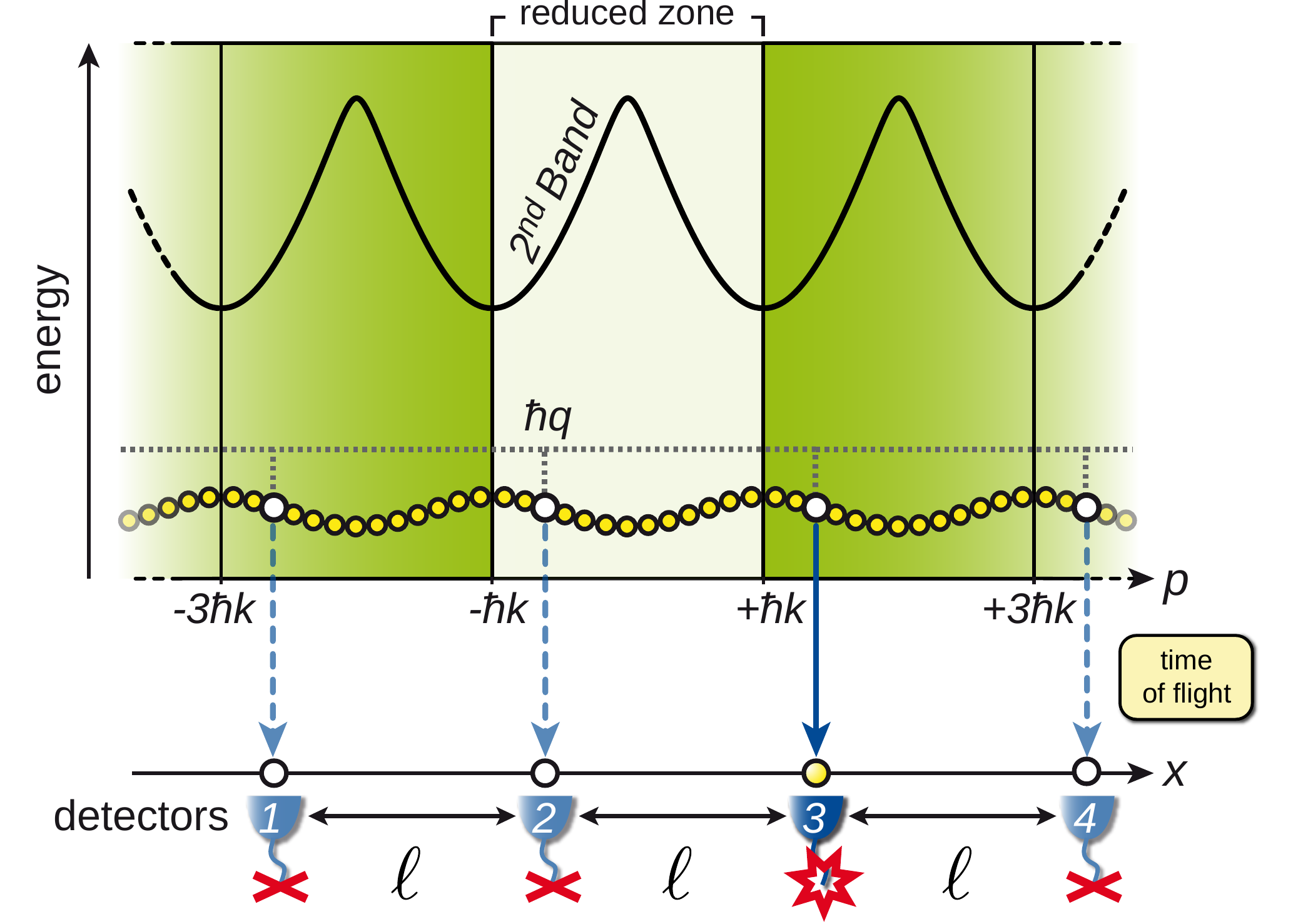}
	\caption{The anticorrelation of fermions released from an optical lattice can be illustrated as a consequence of the Pauli principle. Only one Bloch state can be occupied at a time, leading to only one detection event for all momenta $p$ which correspond to a common lattice quasimomentum $\hbar q$ (figure adapted from \cite{rom06}).}
	\label{fig:sf_FermionSchematics}
\end{figure}

\begin{figure}
	\centering
		\includegraphics[width=1.0\textwidth]{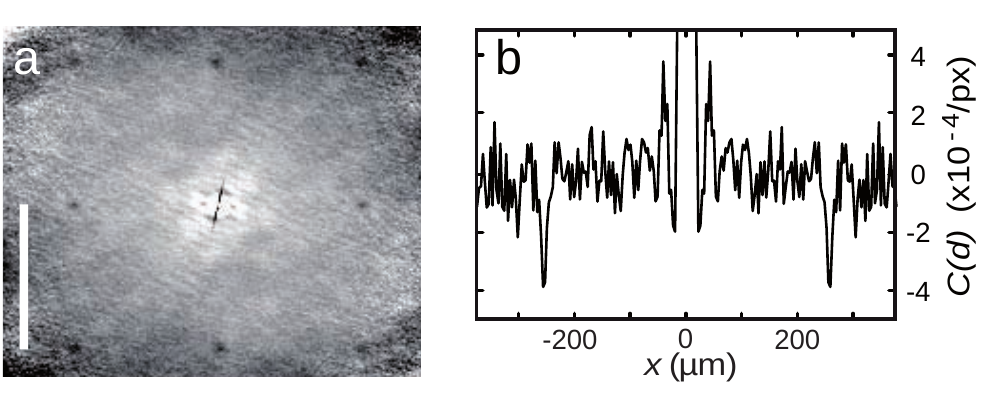}
	\caption{Anticorrelation signal of polarized fermions released from an optical lattice. The autocorrelation peak in the center masks the local antibunching, but anticorrelations at the locations corresponding to the reciprocal lattice positions (white bar denotes $2\hbar k_{lat}$ lattice momentum scale), are clearly visible and negative (figure adapted from \cite{rom06}).}
	\label{fig:sf_FermionCorrelation}
\end{figure}

In the Hanbury-Brown and Twiss picture as well as in the second quantized picture this result is mathematically obvious as a consequence of the sign change, but it is instructive to consider alternative interpretations of the correlation effects. The original proposal by Hanbury-Brown and Twiss for employing density or intensity correlations predicted the appearance of spatial correlations in classical electromagnetic wave signals, without any quantized treatment whatsoever\cite{hanburybrown56a,hanburybrown56b}. So indeed, for the bosonic case, the same results can be obtained using just classical fields. However, for Fermions an analog picture does not work, as the fermionic state can not be directly related to classical fields. One can, however, find another picture for the appearance of the anticorrelations based directly on the Pauli principle. This is illustrated in figure \ref{fig:sf_FermionSchematics}: For unit filling of the optical lattice, the fermionic ensemble is in the band insulating state. This means that every quasimomentum state $\ket q$ within the lowest band is occupied. When releasing this ensemble from the lattice, the Bloch functions corresponding to each $q$ vector are projected to states in free space, which then propagate. However, each Bloch function is constructed only from a subset of all free-space wave functions $\ket k$ such that all involved free-space momenta $p_{q,m}=\hbar k_{q,m}$ can be expressed as $\hbar k_{q,m}=\hbar q+m\cdot 2\hbar k_{lat}$, with $m\in Z$. Therefore, as there can be only one atom for a given $q$, there is a full anticorrelation between its momentum $p$ and all other momenta $p_{q,m}$ corresponding to the same $q$. This leads directly to a periodic correlation with the same structure, but opposite sign, when compared to the bosonic case. Such a signal is shown in figure \ref{fig:sf_FermionCorrelation}.

In the case of polarized fermions, the site occupancy for a degenerate gas (in the lowest band) of a lattice is limited to $n=1$ because of the Pauli principle. One therefore expects a clean $1/N$ scaling of the correlation signal as discussed in section \ref{sec:sf_CorrelAmplitudeDiscussion}. However, thermal excitations in the finite temperature Fermi gases lead to an effective occupation lower than 1, and decrease the correlation amplitude for a given $N$. The pronounced temperature dependence is strongly visible in experiments, and has been proposed as a method to measure the temperature inside the lattice \cite{rom06}.

Using the detection method described in section \ref{sec:sf_3DMetastableCorrelation}, the opposite behavior in the local quantum correlation, bunching vs. anti-bunching, for bosonic and fermionic metastable Helium has also been observed. Here, it was possible to directly observe the local bunching for thermal samples of $^4$He$^*$ as well as the antibunching observed with fermionic $^3$He$^*$, using the very same apparatus for a direct comparison\cite{jeltes07}.

\subsection{Detection of nontrivial spatial order in the lattice\label{sec:sf_PatternedLoading}}

\begin{figure}
		\includegraphics[width=1.0\textwidth]{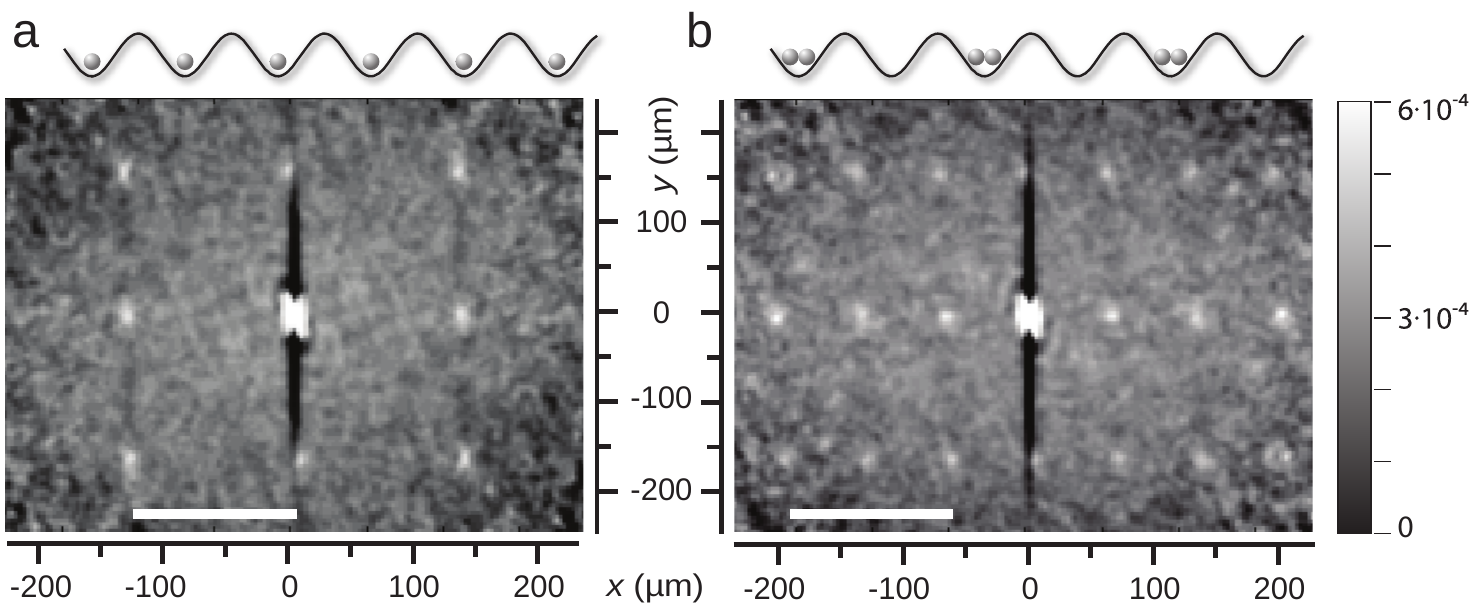}
	\caption{Detection of spatial density modulation inside the lattice. In (a), the correlation pattern from a 3D Mott insulator is shown with all sites occupied, but different lattice constants along the $x$ and $y$ directions of 383\,nm and 420\,nm, respectively. The white bars denote the reciprocal momentum scale for the 383\,nm spaced lattice. If only every second site of the otherwise identical lattice is loaded (b), additional peaks appear in the 2D correlation function, indicating that the density inside the trap is modulated with a period of two lattice sites, along the horizontal axis. The dark vertical lines in the center of the image are an artifact of the imaging caused by filtering against stray light interference. (figure adapted from \cite{foellingPhD})}
	\label{fig:sf_DensityWaveCorrelations}
\end{figure}
Noise correlation analysis with optical lattices is primarily aimed at identifying spatial correlations in the density distribution of the gas, or in the spatial distribution of individual components (such as spin) of the ensemble\cite{demler02,goral02}. As the method works primarily for periodic structures, phases with periodic ordering and density modulations, such as density waves or periodic magnetic order, are specifically suited for correlation analysis. 

The correlations of a simple lattice which is homogeneously filled show the trivial periodicity of the underlying lattice, but any additional order with a periodic structure will show a signal at the corresponding reciprocal momenta. The most simple structures are density waves which are commensurate with the optical lattice, such as shown in figure \ref{fig:sf_DensityWaveCorrelations}. In this example, the two lattice axes are not the same. One is a regular sinusoidal lattice with $420$\,nm periodicity, whereas on the other axis it has a periodicity of 382.5\,nm. In addition, along this axis a second potential with a periodicity of 765\,nm can be applied. This potential can be used during the loading sequence such that every second site can be left empty, for otherwise identical average density. The periodicity of the density distribution has therefore been changed, as a ``density wave'' with a wavelength of two lattice sites has been formed. As a consequence, the correlation pattern now corresponds to that of a 765\,nm periodic lattice structure, and therefore additional correlation peaks at half the original lattice momenta appear.

\begin{figure}
	\centering
		\includegraphics[width=1.0\textwidth]{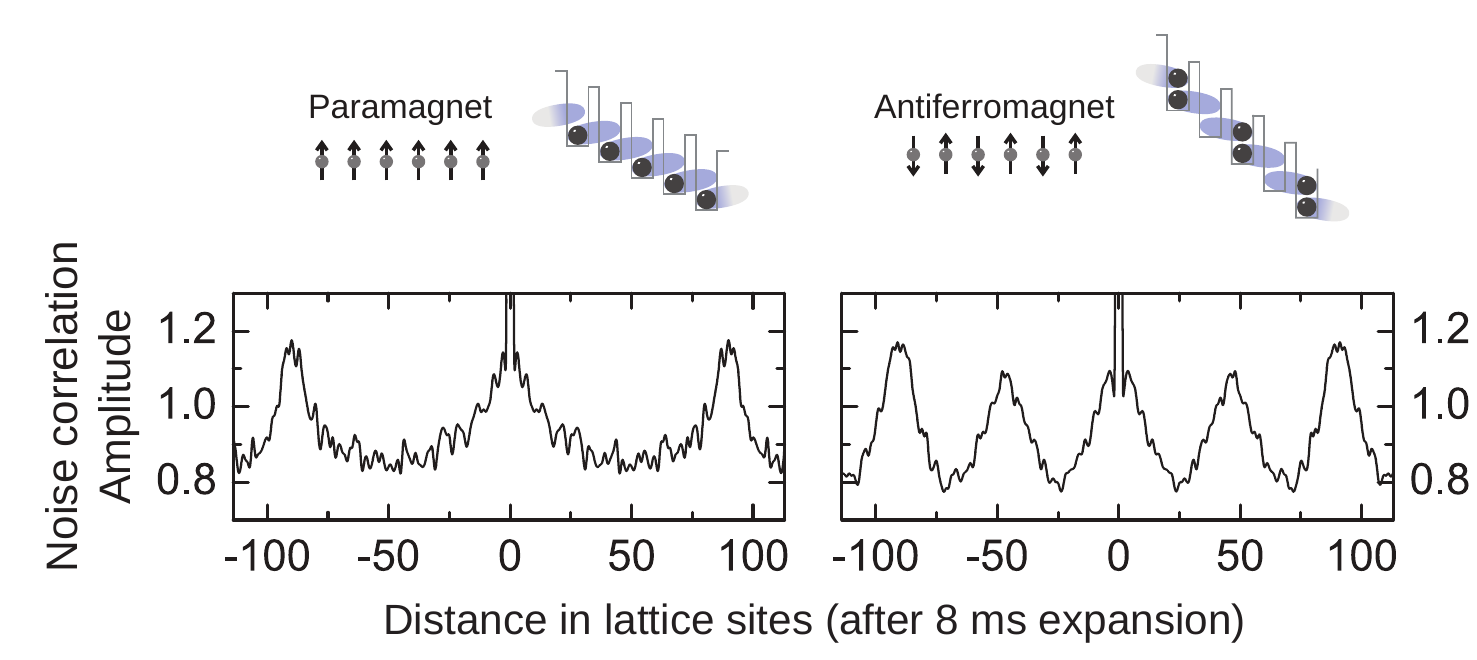}
	\caption{Noise correlation analysis in 1D, comparing a pattern of alternating filling factors and a uniformly filled chain of sites, at very low total density. The initial length of the chain is typically 12 sites and 12 atoms; after expansion by a factor of more than 10, the density is below 0.1 per bin, leading to strong correlations. In the antiferromagnetic case, the additional peaks indicating the increased periodicity are clearly visible. (Figure adapted from\cite{simon2011a}, with permission)}
	\label{fig:sf_Signal_MicroscopeAF}
\end{figure}

More recently, alternating filling of lattice sites has also been achieved by implementing an effective spin chain mapped to the occupation of lattice sites in the optical lattice\cite{simon2011a}. The mapping used allows for the realization of both ferromagnetic and antiferromagnetic states, where the ferromagnet corresponds to unitiy filling of the chain, and the Neel order of the antiferromagnet to a pattern of alternating sites with filling one and two as illustrated in figure \ref{fig:sf_Signal_MicroscopeAF}.
This experiment was conducted in the Harvard quantum gas microscope experiment which uses an imaging system described in chapter \ref{chapterWeitenberg}. This in principle enabled direct in situ density measurements, but a limitation in the imaging process at the time  meant that sites with odd and even occupation could not be distinguished (this effect is described in detail in section \ref{sec:Weitenberg_ParityProjection}). The periodic structure of the effective antiferromagnetic spin chain was therefore observed using noise correlation analysis after the string of atoms was allowed to ballistically expand for 8\,ms along the direction of the chain. The corresponding low atom number per bin (below 1) and the short length of the chain leads to a very high measured correlation amplitude on the order of 0.3, as discussed previously.

\subsection{Momentum correlations from dissociating pairs}

Noise correlation techniques are not limited to systems with optical lattices, but are employed in several circumstances where pair correlations between individual atoms exist, and the density is low enough that these pairs leave a measurable trace in the shot noise. 
\begin{figure}
	\centering
		\includegraphics[width=1.0\textwidth]{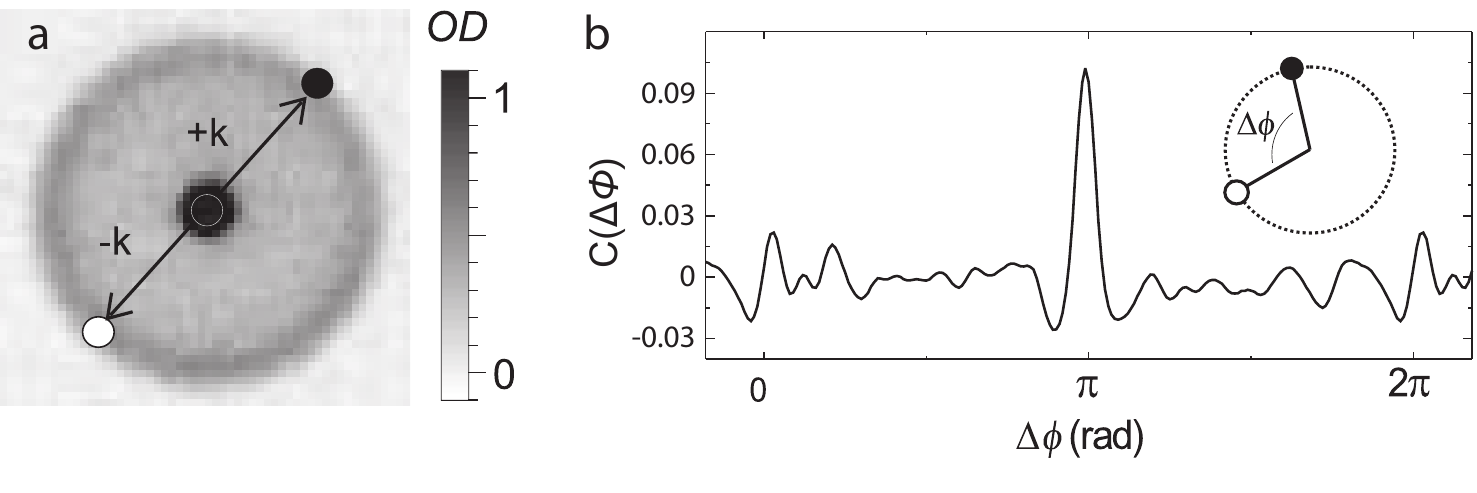}
	\caption{When bound atom pairs at rest are dissociated with finite kinetic energy, the free atoms will fly in opposite directions with a fixed velocity given by the released energy. After ballistic expansion, they therefore appear as a ring around the trap position (left image). Correlating the shot noise along this ring clearly shows the pair relations as a positive correlation for a relative angle of $180\deg$. (Figure adapted from\cite{greiner05}, with permission)}
	\label{fig:sf_DissociationCorrelation}
\end{figure}
One case that leads to strong correlations is that of the dissociation of bound atom pairs with a finite energy. The atoms from the pair are emitted into a random direction, but the center of mass of the pair is unchanged, hence for pairs initially at rest the atoms fly in exactly opposite directions. Such correlations are therefore classical rather than caused by the quantum statistics of the particles, and can therefore also occur with distinguishable particles. They have been detected in  atom pairs released by dissociation of loosely bound Feshbach molecules of degenerate $^{40}$K atoms\cite{greiner05}. The dissociated atoms leave the source region with a velocity which is determined by the dissociation energy, and therefore form a sphere similar to a collision sphere around the center, as shown in figure \ref{fig:sf_DissociationCorrelation}. For correlated pairs, one expects the number fluctuations in this ring to be positively correlated on opposite sides, and this is indeed the case, as shown in the plot which displays the correlation amplitude as a function of relative angle of emission. A peak is found at $\Delta\phi=\pi$, and the signal is flat otherwise. In fact, there is not even an autocorrelation peak at $\Delta\phi=0$, which is a result of the spin-dependent imaging procedure: The pairs always consist of two different spin states $\ket 1$ and $\ket 2$, and these are imaged in two separate images. The correlation analyzed is the inter-state correlation operator corresponding to an expectation value $\langle \opn_1(\phi)\opn_2(\phi+\Delta\phi) \rangle$. As this is no autocorrelation, the central peak does not appear.

\subsection{Pair correlations caused by collisions}
Momentum correlations between atom pairs are also induced when atoms with well-defined relative momenta collide, as the center of mass-momentum remains unchanged. These correlations can then become visible in the number fluctuations of the atoms exiting the collision.
\begin{figure}
	\centering
		\includegraphics[width=1.0\textwidth]{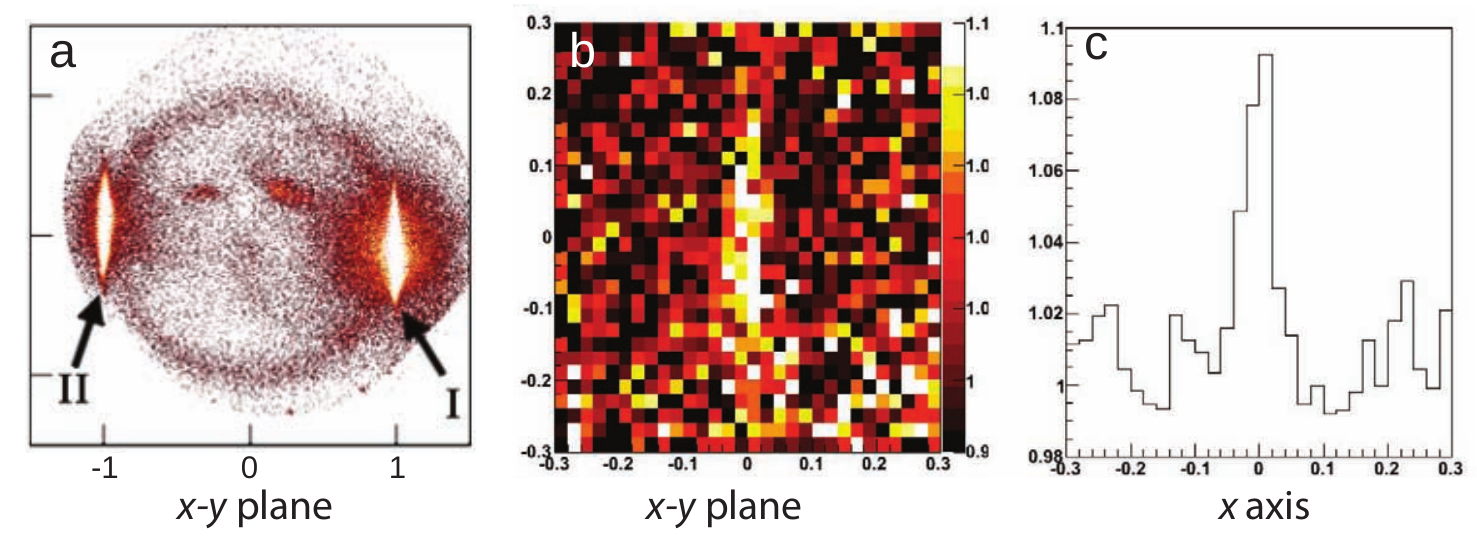}
	\caption{Correlated atom pairs after a collision of two BECs with defined momenta. Figure a shows a slice of the 3D momentum distribution approximately 320\,ms after the collision. The two remaining BECs leaving the collision point can can be seen on the left and the right, with the scattered atoms forming an $s$-wave collision sphere. The scattered atoms are correlated for opposite momenta, as shown in the density-density correlation function (b). Here, 0 denotes to exactly opposite momenta, corresponding to zero center of mass momentum. (c) shows a profile through the center of (b), with a detected amplitude of almost 1.1. (Figure adapted from \cite{Perrin2007} and \cite{Westbrook2006}, with permission)}
	\label{fig:sf_CollisionCorrelations}
\end{figure}
In one example of such an experiment, the full 3D correlation function of such atom pairs has been measured, by using the type of setup described in section \ref{sec:sf_3DMetastableCorrelation}. To induce well-defined collisions, a BEC was split into two parts using a Raman laser process which imparted a well-defined momentum to one half of the BEC and the opposite momentum to the other half. These two moving halves of the condensate then collide, resulting in coherent (four-wave mixing) and non-coherent scattering of atoms. The incoherent scattering part is solely due to simple two-body collisions, hence the center of mass momentum of the pair stays zero, and the particles are ejected into opposite directions and form an s-wave collision sphere. This sphere can be seen in figure \ref{fig:sf_CollisionCorrelations}(a), as well as the remainder of the two parts of the BEC, which are also moving outward from collision center with the same velocity. Correlating opposite sides of the collision sphere clearly shows the correlated momentum pairs, with a correlation amplitude only limited by the quantum efficiency of the detector and the size of the collision region. It should be noted that, while the collision process is a classical process generating correlations within the colliding pair, quantum correlations (bunching) do occur in addition, between atoms from separate pairs\cite{Molmer2008}. This leads to correlations also in copropagating atoms, which are indeed detected in the experiment\cite{Westbrook2006}.

A natural way of interpreting a positive correlation of densities in two locations scenario is to see it as a suppression of density differences. Integrating the total atom numbers over similarly shaped volumes on opposite sides of the collision center should therefore show atom number differences which are below those  expected for a completely random distribution of atoms (a negative correlation would lead to an increase in the noise). Such reduced fluctuations can in fact be detected by suitable integration in He$^*$ experiments\cite{Jaskula2010}.

At the opposite end of the detection methods compared to the full 3D discrimination is the extreme case of a noise correlation measurement where only two detection regions are being used and ideally only two modes are correlated with one another. If atom pairs in both regions are positively correlated, atom number differences are suppressed, and the two modes are in a number-squeezed state. The detection of the effect of course requires a sub shot noise sensitive atom number counting method. Typically, this will of course work best with relatively low atom numbers (for high relative shot noise), and high-sensitivity imaging. 

Such reduced relative atom number fluctuations have been created in a two-mode system by colliding two BECs in a similar way as described above, but with the scattered atoms being confined to a waveguide-type potential on an atom chip, allowing for only two (opposite)scattering directions \cite{Brueckner2011}. Counting the number in each of these two modes by integrating the densities with standard optical imaging revealed reduced relative fluctuation and thus positive correlation of the two modes.

\section{Conclusion}
We have introduced noise correlation analysis as a tool to analyze the properties of many-body states as well as dynamic processes involving interacting atoms. For optical lattices with strongly correlated ensembles, the intrinsic quantum correlations in the noise allow for the determination of spatial structures inside the lattice, even in time of flight measurements after the ensemble has been released from the original configuration. They can therefore serve as a detection tool for quantum states in lattices with nontrivial spatial structure, even when no in-situ imaging method with the required resolution is available or possible. However, quantum noise correlations are also a fascinating and fairly simple demonstration of the quantum character of the particle ensembles, and its effect on both bosonic and fermionic quantum gases.


\setcounter{tocdepth}{4}
\end{document}

%% file: commands.tex
\newcommand{\opn}{{\hat n}}

\newcommand{\opb}{{\hat b}}
\newcommand{\opbd}{{\hat b^\dagger}}

\newcommand{\vect}[1]{\mathbf{#1}}

\newcommand{\average}[1]{\langle#1\rangle}
\newcommand{\bigaverage}[1]{\left\langle#1\right\rangle}

\newcommand{\ket}[1]{\lvert#1\rangle}

\newcommand{\ee}{\text{e}}

